\documentclass[final,5p,times,twocolumn]{elsarticle}

\usepackage{graphicx}
\usepackage{amssymb}
\usepackage{amsmath}
\usepackage{enumitem}
\usepackage{multirow}
\usepackage{tikz}
\usepackage{epstopdf}
\usepackage[percent]{overpic}
\usepackage{scrextend}
\usepackage{xcolor,xspace}
\usepackage[ulem=normalem]{changes}
\usepackage{hyperref}
\hypersetup{colorlinks=True,urlcolor=blue,linkcolor=blue,citecolor=blue,filecolor=black}
\usepackage{dcolumn,color,footnote,bm,braket}
\usepackage{threeparttable}
\usepackage{fancybox}

\usetikzlibrary{matrix}

\colorlet{Changes@Color}{red}

\newcommand\+{\dagger}

\newcommand\mzn{M^{0\nu}}
\newcommand\mgt{M_{\mathrm{GT}}^{0\nu}}
\newcommand\mfe{M_{\mathrm{F}}^{0\nu}}
\newcommand\mte{M_{\mathrm{T}}^{0\nu}}

\newcommand\znbb{0\nu\beta\beta}
\newcommand\ga{g_{\mathrm{A}}}
\newcommand\gv{g_{\mathrm{V}}}

\newcommand\intr{[N_{\nu}\otimes N_{\pi}']}
\newcommand\norm{[N_{\nu}\otimes N_{\pi}]}

\journal{Physics Letters B}

\begin{document}

\begin{frontmatter}

\title{
Configuration mixing effects
on neutrinoless 
$\beta\beta$-decay nuclear matrix elements}

\author[1,2]{Kosuke Nomura\corref{cor}}
\ead{nomura@sci.hokudai.ac.jp}

\affiliation[1]
{
organization={Department of Physics, Hokkaido University},
city={Sapporo},
postcode={060-0810},
country={Japan}
}

\affiliation[2]
{
organization={Nuclear Reaction Data Center, Hokkaido University},
city={Sapporo},
postcode={060-0810},
country={Japan}
}

\cortext[cor]{Corresponding author}

\begin{abstract}
Mixing and coexistence of intrinsic nuclear shapes
play an important role to determine
the low-energy structure of heavy nuclei, and
are expected to affect
nuclear matrix elements (NMEs) of
neutrinoless double beta ($0\nu\beta\beta$) decay.
This problem is addressed in the interacting
boson model with
configuration mixing
that is formulated by using
the nuclear energy density functional theory.
It is shown that significant amounts of
mixing of normal and deformed intruder
configurations are present in the ground
and excited $0^+$ states in the even-even nuclei
that are parent or daughter nuclei of
the $0\nu\beta\beta$ decay.
An illustrative application to
the $0\nu\beta\beta$ decays of $^{76}$Ge, $^{96}$Zr,
$^{100}$Mo, $^{116}$Cd, and $^{150}$Nd shows
that the inclusion of the
configuration mixing reduces the NMEs
for most of the $0^+_1$ $\to$ $0^+_1$
$0\nu\beta\beta$ decays.
\end{abstract}

\date{\today}


\end{frontmatter}

Neutrinoless double beta ($\znbb$) decay
is of broad physical significance,
as its observation would provide
insights into beyond-standard-model physics,
and masses and nature of neutrinos.
Related experiments are
ongoing and planned worldwide.
Theoretical investigations
for the $\znbb$ decay have mainly
concerned accurate predictions of
the nuclear matrix element (NME)
using various nuclear many-body methods
\cite{avignone2008,engel2017,agostini2023,gomezcadenas2024}.
These NME predictions,
however, differ by several factors,
necessitating further sophistication
of a given theoretical method
by taking into account various
important nuclear correlations.

A distinct feature of the
atomic nucleus
is the deformation of its surface,
leading to a variety of intrinsic
shapes and collective excitations.
Multiple shapes can often
coexist in a single nucleus.
The phenomenon of shape coexistence
has been observed in many nuclides,
and the precise description of
this phenomenon attracts considerable
interests in nuclear structure studies
\cite{wood1992,heyde2011,garrett2022,leoni2024}.
An empirical signature of
the shape coexistence is the
appearance of
low-energy $0^+$ excited states,
which can be interpreted to be
intruder states arising from 
particle-hole excitations
\cite{heyde85,heyde87,federman77}.
The shape coexistence is also inferred
from the appearance of several energy minima
in the potential energy surface
(PES) in mean-field models
\cite{bengtsson1987,cwiok2005,andre00}.
The coexistence and mixing
of multiple shapes
play pivotal parts in the
low-lying states in nuclei, and
could have potential impacts
on the $\znbb$-decay NMEs.

The interacting boson model (IBM) \cite{IBM},
in which collective pairs of
valence nucleons are represented by
bosons, has allowed
for detailed and systematic analyses of the
collective excitation spectra and
transition rates of many nuclei,
and has also been extensively used
for the $\znbb$-decay studies
\cite{barea2009,barea2013,barea2015,deppisch2020}.
In these IBM calculations,
the model parameters
for the even-even nuclei
were those determined from experimental
low-energy spectra.
The Gamow-Teller (GT), Fermi,
and tensor transition 
operators were formulated by using
a generalized seniority scheme.
The predicted $\znbb$-decay NMEs
were shown to be typically
close to those from
the quasiparticle
random-phase approximation
(QRPA) and generator coordinate method (GCM)
but to be much larger than
the nuclear shell model (NSM) values.
More recently,
another set of the IBM calculations
for the $\znbb$ decay
that is formulated by using the
nuclear energy density
functional (EDF) approach
was presented \cite{nomura2025bb}.
In that study,
the strength parameters for
the IBM Hamiltonians for the
parent and daughter nuclei
are determined by mapping the
PES, computed by the self-consistent
mean-field (SCMF) method
employing a given EDF,
onto the equivalent energy surface
in the boson system,
for which no phenomenological
adjustment to experimental
data is needed.
The $\znbb$-decay operators were
constructed by exploiting the method of
\cite{barea2009}.
The resultant NMEs were
shown to be smaller than those
in these earlier IBM-2 values \cite{nomura2025bb}.
In all these IBM predictions,
however, intruder configurations have
never been included,
which are relevant to
the nature of the ground and excited
$0^+$ states.

This article addresses
the shape coexistence and mixing effects
on the $\znbb$-decay NMEs
using the IBM with configuration mixing (IBMCM).
This is based on the preceding
study of \cite{nomura2025bb},
but the mixing of multiple
shape configurations is here explicitly
taken into account for the calculations
of even-even nuclei, and in
the formulations of the transition
operators.
Updated IBM predictions are
given for the
$\znbb$-decay NMEs of the
candidate nuclei $^{76}$Ge,
$^{96}$Zr, $^{100}$Mo, $^{116}$Cd,
and $^{150}$Nd,
and the relevance of the
configuration mixing in these decay
processes is investigated.

The IBMCM is an extension of the IBM
to include contributions
from outside of the normal
(a given valence) space \cite{duval1981}.
These additional configurations are associated 
with intruder (and usually deformed) states
analogous to those in the shell model,
and the normal and intruder configurations
are mixed.
The intruder configurations
refer to two-particle-two-hole ($2p$-$2h$), 
four-particle-four-hole ($4p$-$4h$), $\ldots$ 
excitations of neutrons or protons,
which are described by 
the IBM comprising
$N_{\rm B}+2$, $N_{\rm B}+4$, $\ldots$ bosons,
provided the like-hole bosons are
not distinguished from the
like-particle bosons.
Here $N_{\rm B}$ refers to the
total number of bosons in
the normal IBM space.

In the following,
the neutron-proton IBM (IBM-2) is
used, in which a given even-even
nucleus is described as a system of neutron 
$s_{\nu}$ and $d_{\nu}$ bosons and 
proton $s_{\pi}$ and $d_{\pi}$ bosons
\cite{OAIT,OAI,IBM}.
$s_{\nu}$ and $d_{\nu}$ ($s_{\pi}$ and $d_{\pi}$) 
bosons reflect collective
monopole and quadrupole 
pairs of valence neutrons (protons), respectively. 
The neutron (proton)
$N_{\nu}$ ($N_\pi$) boson
number is equal to half the number
of the neutron (proton) pairs,
and $N_{\rm B}=N_{\nu}+N_{\pi}$
is conserved for a given nucleus.
Here, particle-hole excitations 
are assumed to be those of protons,
and the intruder configurations
up to $2p$-$2h$  excitations are considered. 
The IBMCM configuration space
is expressed as
a direct sum of subspaces representing
the normal $\norm$ and intruder
$[N_{\nu}\otimes (N_{\pi}+2)]$ spaces:
\begin{eqnarray}
\label{eq:ibmcmspace}
 [N_{\nu}\otimes N_{\pi}]\oplus
[N_{\nu}\otimes (N_{\pi}+2)] \; .
\end{eqnarray}
Using a short-hand notation
$N_{\pi}'\equiv N_{\pi}+2$,
the IBMCM Hamiltonian consists of
two independent Hamiltonians for
the $\norm$ and $\intr$ spaces that
differ in boson number by 2:
\begin{eqnarray}
\label{eq:ibmcmham}
 \hat H = \hat P_{N_{\pi}} \hat H_{N_{\pi}} \hat P_{N_{\pi}}
+ \hat P_{N_{\pi}'} ( \hat H_{N_{\pi}'} +\Delta ) \hat P_{N_{\pi}'}
+ \hat V_{\rm mix} \; ,
\end{eqnarray}
where $\hat P_{N_{\pi}}$ and $\hat P_{N_{\pi}'}$ 
are projection operators onto the 
$[N_{\nu}\otimes N_{\pi}]$ and 
$[N_{\nu}\otimes N_{\pi}']$ spaces, respectively, 
$\hat H_{N_{\pi}}$ and $\hat H_{N_{\pi}'}$ 
are the unperturbed Hamiltonians, 
$\Delta$ represents an offset energy 
required for the particle-hole excitations, 
and $\hat V_{\rm mix}$ is the interaction 
that mixes the two boson subspaces.
The forms of the unperturbed
IBM-2 Hamiltonian and mixing interaction
$\hat V_{\rm mix}$ are
adopted from \cite{nomura2025bb}
and \cite{nomura2013hg},
respectively.

To construct the IBMCM Hamiltonian,
constrained SCMF calculations of
the PESs are performed
within
the relativistic Hartree-Bogoliubov
(RHB) method
\cite{vretenar2005,niksic2011,DIRHB,DIRHBspeedup}
using the density-dependent
point-coupling (DD-PC1) interaction \cite{DDPC1}
and separable pairing force of
finite range \cite{tian2009},
and within the Hartree-Fock-Bogoliubov
(HFB) \cite{robledo2019,hfbtho400} method using
the D1M \cite{D1M} interaction of the
nonrelativistic Gogny EDF \cite{gogny}.
These two interactions are
representative classes
of the nuclear EDF, i.e., relativistic
and nonrelativistic,
and the two sets of the SCMF calculations,
denoted hereafter as RHB and HFB,
are here performed to
show robustness of the mapped IBMCM.

One can infer geometry
of a given IBM Hamiltonian
by taking its expectation value
in the boson coherent state
\cite{ginocchio1980,dieperink1980,bohr1980},
which results in an energy surface
in terms of the
deformation variables
analogous to the axial quadrupole
deformation $\beta$ and triaxiality
$\gamma$ in the geometrical model \cite{BM}.
The energy surface for the IBMCM
is obtained as an eigenvalue
of a $2\times 2$ coherent-state
matrix \cite{frank2004}.
A procedure to fix the
IBMCM Hamiltonian is such that
\cite{nomura2012sc}
the unperturbed Hamiltonian
for the normal configuration
is determined
so that the expectation value of
the Hamiltonian should 
reproduce topology of the SCMF PES near 
a minimum corresponding to 
the smallest $\beta$ deformation, 
and that the unperturbed
Hamiltonian for the
intruder configuration is
associated with the mean-field
minimum with 
a larger $\beta$ deformation. 
The offset $\Delta$ and mixing strength
are determined 
so that the lowest eigenvalue of the 
coherent-state matrix should reproduce 
the energy difference between
the two mean-field minima and the
topology of the PES
between these minima.
The resulting IBMCM Hamiltonian
is diagonalized
in the space \eqref{eq:ibmcmspace},
providing the energies and
wave functions for even-even nuclei.

The NME of the $\znbb$
decay $\mzn$ consists of 
the Gamow-Teller (GT), Fermi (F), 
and tensor (T) components:
\begin{eqnarray}
 \mzn = \mgt - \left(\frac{\gv}{\ga}\right)^2 \mfe + \mte \; ,
\end{eqnarray}
where $\ga=1.269$ and $\gv=1$ \cite{Yao2006} 
are the axial-vector 
and vector coupling constants, respectively. 
The matrix elements of the components in $\mzn$ 
for the $0^+\,\to\,0^+$ $\znbb$ decay are 
computed by using the eigenfunctions of the 
IBMCM Hamiltonian
\eqref{eq:ibmcmham} for the 
initial $\ket{0^+_{\rm I}}$ state in parent 
and final $\ket{0^+_{\rm F}}$ state in daughter nuclei, 
$\mzn_{\alpha} = 
\braket{0^+_{\rm F}| \hat O_\alpha |0^+_{\rm I}}$,
with $\alpha$ = GT, F, and T.
As in previous IBM studies
for the $\znbb$ decay
\cite{barea2009,barea2015,deppisch2020,nomura2025bb},
the NMEs are computed by assuming
the closure approximation,
in which intermediate states
of the neighboring odd-odd nuclei
are not explicitly
taken into account.
The operator $\hat O_\alpha$ is given by
\cite{barea2009}
\begin{align}
\label{eq:mzn-1}
 \hat O_\alpha
= 
&- \frac{1}{2}
\sum_{j_{\pi}}
\sum_{j_{\nu}}
O_{\alpha}(j_{\pi}j_{\pi}j_{\nu}j_{\nu};0)
\nonumber\\
&\quad
\times
A^{(+)}_\pi(j_{\pi};N_{\pi,{\rm I}})
A^{(-)}_{\nu}(j_{\nu};N_{\nu,{\rm I}})
s^\+_\pi\cdot\tilde s_{\nu}
\nonumber\\
&- \frac{1}{4}
\sum_{j_{\pi 1}j_{\pi 2}}
\sum_{j_{\nu 1}j_{\nu 2}}
\sqrt{1+(-1)^J\delta_{j_{\pi 1}j_{\pi 2}}}
\sqrt{1+(-1)^J\delta_{j_{\nu 1}j_{\nu 2}}}
\nonumber \\
&\quad
\times
O_{\alpha}(j_{\pi 1}j_{\pi 2}j_{\nu 1}j_{\nu 2};2)
\nonumber\\
&\quad
\times
B^{(+)}_\pi(j_{\pi 1}j_{\pi 2};N_{\pi,{\rm I}})
B^{(-)}_{\nu}(j_{\nu 1}j_{\nu 2};N_{\nu,{\rm I}})
d^\+_\pi \cdot \tilde d_{\nu}
\; ,
\end{align}
where the first and second terms
represent the monopole ($J=0^{+}$)
and quadrupole ($J=2^{+}$)
contributions, respectively.
$A^{(\pm)}_\rho(j_{\rho};N_{\rho,{\rm I}})$ and 
$B^{(\pm)}_\rho(j_{\rho 1}j_{\rho 2};N_{\rho,{\rm I}})$ 
($\rho=\pi,\nu$)
denote coefficients 
for the one-$s_{\rho}$ and one-$d_{\rho}$ boson 
transfer [addition $(+)$ and removal $(-)$] 
operators, respectively,
and are calculated within
the generalized seniority scheme
\cite{frank1982,barea2009}.
These coefficients involve
pair structure constants for the
collective monopole and quadrupole
pairs, which are computed by using
the occupation probabilities for
the relevant single-particle orbits
provided by the EDF-SCMF
calculations \cite{nomura2025bb}.
Note that $N_{\rho,{\rm I}}$ and $N_{\rho,{\rm F}}$ 
are boson numbers in the parent and daughter nuclei, 
respectively. 
$N_{\nu,{\rm F}}=N_{\nu,{\rm I}} \mp 1$ for 
like-particle ($-$) and like-hole ($+$) neutron bosons, 
while 
$N_{\pi,{\rm F}}=N_{\pi,{\rm I}} \pm 1$ 
for like-particle ($+$)
and like-hole ($-$) proton bosons. 
The coefficients
$A^{(+)}_{\rho}$ and $A^{(-)}_{\rho}$ are related by
$A^{(-)}_{\rho}(j_{\rho};N_{\rho})=-A^{(+)}_{\rho}(j_{\rho};N_{\rho}-1)$, 
while 
$B^{(-)}_{\rho}(j_{\rho 1}j_{\rho 2};N_{\rho})=(-1)^{j_{\rho 1}+j_{\rho 2}}B^{(+)}_{\rho}(j_{\rho 1}j_{\rho 2};N_{\rho}-1)$. 
If neutron (proton) bosons are treated as holes, 
the neutron annihilation (proton creation) operators 
in (\ref{eq:mzn-1}) should be replaced 
with the creation (annihilation) operators. 
The forms for these coefficients 
are found in
Appendix A 3 of Ref.~\cite{nomura2025bb}. 
The fermion two-body matrix element
$O_{\alpha}(j_{\pi 1}j_{\pi 2}j_{\nu 1}j_{\nu 2};J)$ 
in (\ref{eq:mzn-1})
depends on the neutrino potential, and
is calculated with the method
described in Refs.~\cite{barea2009,nomura2025bb}.
In addition, the short-range correlation
is taken into account for the radial part of the 
potential by using 
the Jastrow function squared
with the Argonne v18 parametrization for
the $NN$ force \cite{simkovic2009}.

As in the case of the standard
IBM-2 with a single configuration,
it is assumed that transfers
of more than one proton bosons
are negligible in the IBMCM.
Under this assumption, three types
of the transition are possible:
normal-to-normal,
$[N_{\nu,{\rm I}}\otimes N_{\pi,{\rm I}}]
\to
[N_{\nu,{\rm F}}\otimes N_{\pi,{\rm F}}]$,
intruder-to-intruder,
$[N_{\nu,{\rm I}}\otimes N_{\pi,{\rm I}}']
\to
[N_{\nu,{\rm F}}\otimes N_{\pi,{\rm F}}']$
and, if proton bosons are treated
as particles (or holes), intruder-to-normal,
$[N_{\nu,{\rm I}}\otimes N_{\pi,{\rm I}}']
\to
[N_{\nu,{\rm F}}\otimes N_{\pi,{\rm F}}]$
(or normal-to-intruder,
$[N_{\nu,{\rm I}}\otimes N_{\pi,{\rm I}}]
\to
[N_{\nu,{\rm F}}\otimes N_{\pi,{\rm F}}']$),
configurations.
The operators
$A^{(+)}_{\pi}(j_{\pi};N_{\pi}) s^\+_{\pi}$ and
$B^{(+)}_{\pi}(j_{\pi 1}j_{\pi 2};N_{\pi}) d^\+_{\pi}$ 
in \eqref{eq:mzn-1}
should be extended accordingly as
\begin{align}
 A^{(+)}_{\pi}(N_{\pi,{\rm I}}) s^\+_{\pi}
\rightarrow
&
\quad A^{(+)}_{\pi}(N_{\pi,{\rm I}}) \hat P_{N_{\pi,{\rm F}}} s^\+_{\pi} \hat P_{N_{\pi,{\rm I}}}
+ A^{(+)}_{\pi}(N_{\pi,{\rm I}}') \hat P_{N_{\pi,{\rm F}}'} s^\+_{\pi} \hat P_{N_{\pi,{\rm I}}'}
\nonumber\\
&
+ A^{(-)}_{\pi}(N_{\pi,{\rm I}}') \hat P_{N_{\pi,{\rm F}}} \tilde s_{\pi} \hat P_{N_{\pi,{\rm I}}'}
\end{align}
\begin{align}
 B^{(+)}_{\pi}(N_{\pi,{\rm I}}) d^\+_{\pi}
\rightarrow
&
\quad B^{(+)}_{\pi}(N_{\pi,{\rm I}}) \hat P_{N_{\pi,{\rm F}}} d^\+_{\pi} \hat P_{N_{\pi,{\rm I}}}
+ B^{(+)}_{\pi}(N_{\pi,{\rm I}}') \hat P_{N_{\pi,{\rm F}}'} d^\+_{\pi} \hat P_{N_{\pi,{\rm I}}'}
\nonumber\\
& + B^{(-)}_{\pi}(N_{\pi,{\rm I}}') \hat P_{N_{\pi,{\rm F}}} \tilde d_{\pi} \hat P_{N_{\pi,{\rm I}}'}
\; ,
\end{align}
where the arguments $j_{\pi}$, $j_{\pi 1}$, 
and $j_{\pi 2}$ are
omitted for brevity. 
For like-hole protons, similar expressions
\begin{align}
 A^{(-)}_{\pi}(N_{\pi,{\rm I}}) \tilde s_{\pi}
\rightarrow
&
\quad A^{(-)}_{\pi}(N_{\pi,{\rm I}}) \hat P_{N_{\pi,{\rm F}}} \tilde s_{\pi} \hat P_{N_{\pi,{\rm I}}}
+ A^{(-)}_{\pi}(N_{\pi,{\rm I}}') \hat P_{N_{\pi,{\rm F}}'} \tilde s_{\pi} \hat P_{N_{\pi,{\rm I}}'}
\nonumber\\
&
+ A^{(+)}_{\pi}(N_{\pi,{\rm I}}') \hat P_{N_{\pi,{\rm F}}'} s^{\+}_{\pi} \hat P_{N_{\pi,{\rm I}}}
\end{align}
\begin{align}
 B^{(-)}_{\pi}(N_{\pi,{\rm I}}) \tilde d_{\pi}
\rightarrow
&
\quad B^{(-)}_{\pi}(N_{\pi,{\rm I}}) \hat P_{N_{\pi,{\rm F}}} \tilde d_{\pi} \hat P_{N_{\pi,{\rm I}}}
+ B^{(-)}_{\pi}(N_{\pi,{\rm I}}') \hat P_{N_{\pi,{\rm F}}'} \tilde d_{\pi} \hat P_{N_{\pi,{\rm I}}'}
\nonumber\\
&
+ B^{(+)}_{\pi}(N_{\pi,{\rm I}}') \hat P_{N_{\pi,{\rm F}}'} d^{\+}_{\pi} \hat P_{N_{\pi,{\rm I}}}
\; ,
\end{align}
are used.

%
%
\begin{figure}[ht]
\begin{center}
\includegraphics[width=\linewidth]{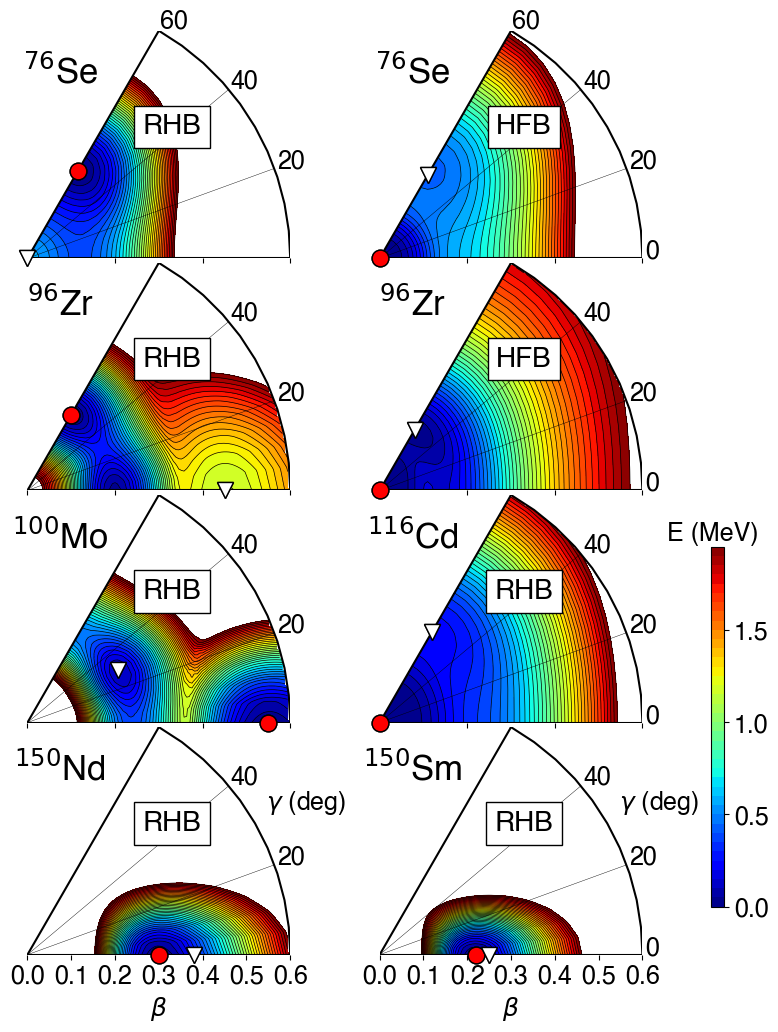}
\caption{
Potential energy surfaces in terms of the 
triaxial quadrupole $(\beta,\gamma)$ deformations 
for the nuclei of interest in the IBMCM
that is based on the RHB and HFB
SCMF calculations.
The global and local minima are indicated
by the solid circles and open triangles,
respectively.}
\label{fig:pes}
\end{center}
\end{figure}

The above procedure is
applied to the $\znbb$ decays
$^{76}$Ge $\to$ $^{76}$Se, 
$^{96}$Zr $\to$ $^{96}$Mo, 
$^{100}$Mo $\to$ $^{100}$Ru, 
$^{116}$Cd $\to$ $^{116}$Sn,
and
$^{150}$Nd $\to$ $^{150}$Sm.
The configuration mixing
is considered for the nuclei
$^{76}$Se, $^{96}$Zr,
$^{100}$Mo, $^{116}$Cd,
$^{116}$Sn, $^{150}$Nd, and $^{150}$Sm.
For $^{116}$Cd in particular,
the IBMCM calculation is made only
with the RHB input,
since there is no local
minimum in the HFB PES
\cite{nomura2025bb}.
Figure~\ref{fig:pes} displays
illustrative examples for
the quadrupole triaxial
$(\beta,\gamma)$ mapped IBMCM PESs.
The RHB PES for $^{76}$Se shows
an oblate global minimum 
at $\beta\approx 0.23$,
and a spherical local minimum.
The HFB PES for the same nucleus
exhibits a spherical ground state
and an oblate local minimum.
The RHB PES for $^{96}$Zr
is soft in $\gamma$ deformation but is 
rather steep in $\beta$ deformation.
Note that a prolate minimum
at $\beta\approx 0.2$ for this
nucleus arises from
a single IBM Hamiltonian
for the normal configuration,
for which a negative strength parameter
for the three-body boson 
interaction is used.
The intruder configuration in this case
is, therefore, associated with the
local prolate minimum
at $\beta\approx 0.45$.
The HFB PES for $^{96}$Zr suggests
a spherical global minimum and
an oblate local minimum,
which are quite close in energy.
The RHB
PES for $^{100}$Mo exhibits a triaxial
minimum at $\gamma\approx 30^{\circ}$,
and a pronounced prolate equilibrium
minimum at $\beta\approx 0.55$.
For $^{116}$Cd,
an oblate local minimum is predicted
in addition to the spherical global minimum
in the RHB PES.

A well developed prolate minimum
is found for $^{150}$Nd
and $^{150}$Sm, but there is
no noticeable local minimum.
The IBMCM calculation is nevertheless
performed for these nuclei,
on the basis of the fact that
the observed $0^+_2$ energy-levels
are substantially low,
and the introduction of the configuration
mixing appears to be necessary.
Shape/phase coexistence 
has indeed been empirically suggested
to occur near $N=90$
\cite{heyde2011,garrett2022}.
For these nuclei,
the Hamiltonian for the intruder
configuration is associated with
the mean-field configuration
corresponding to
the $\beta$ deformation that is
slightly larger than
that of the global minimum,
because at that
larger $\beta$ deformation
the curvature in $\beta$ changes.
These associations of the
IBM configurations are also inspired
by a recent Monte-Carlo Shell-Model
study for the $^{150}$Nd $\znbb$
decay \cite{tsunoda2023},
in which the $0^+_1$ wave functions
for $^{150}$Nd and $^{150}$Sm
are populated at the configurations
corresponding to a prolate
deformation and to a slightly
larger one also on the prolate side.

%
%
\begin{figure}[ht]
\begin{center}
\includegraphics[width=\linewidth]{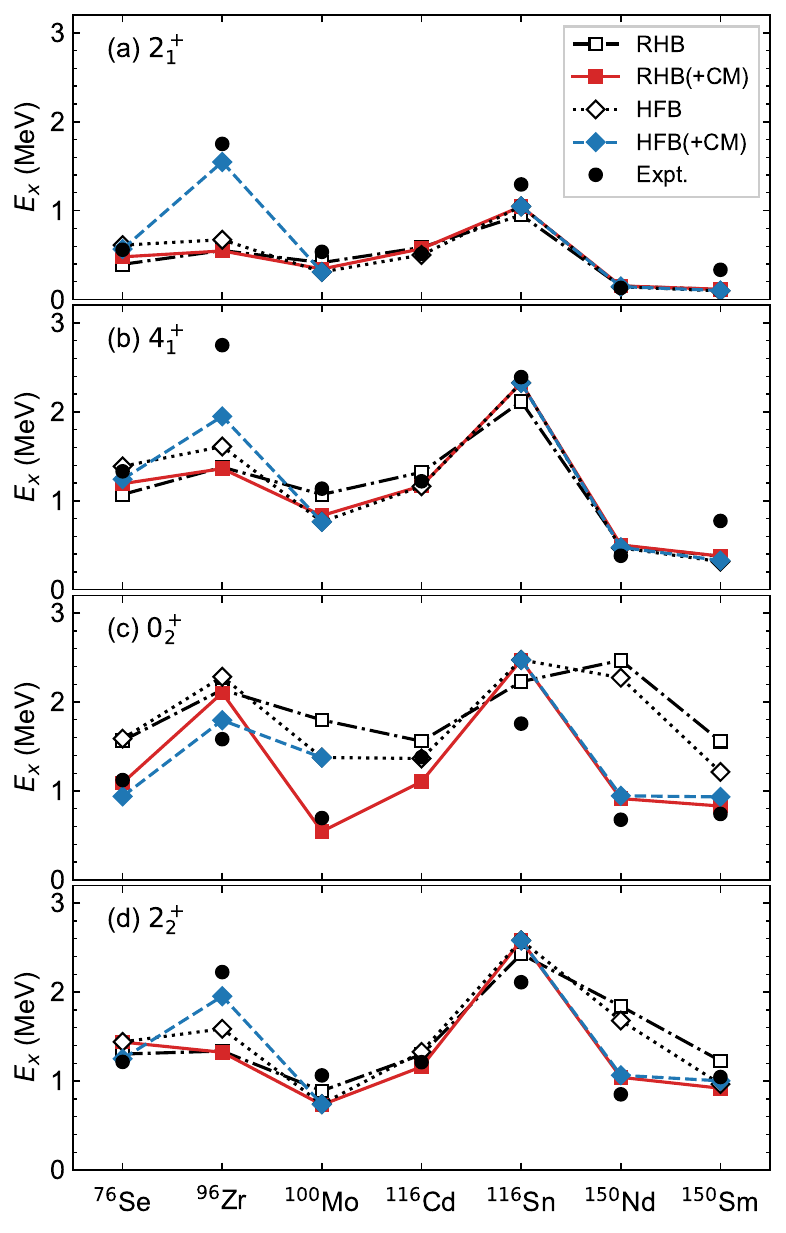}
\caption{
Calculated excitation energies
for the $2^+_1$,
$4^{+}_1$, $0^+_2$, and $2^+_2$ states
in the IBMCM that is based
on the RHB and HFB SCMF methods.
The results obtained from the IBM
without configuration mixing
are adopted from \cite{nomura2025bb}.
Experimental data are taken
from the NNDC database \cite{data}. 
}
\label{fig:energy}
\end{center}
\end{figure}

%
%
\begin{figure}[ht]
\begin{center}
\includegraphics[width=\linewidth]{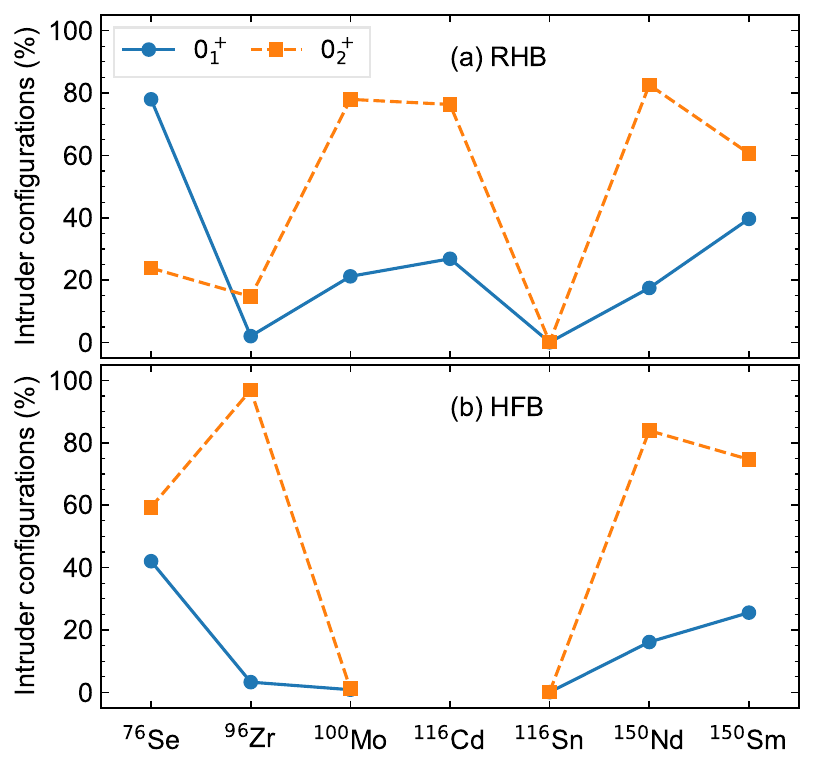}
\caption{
Fractions of the intruder configurations
in the IBMCM wave functions of the 
$0^+_1$ and $0^+_2$ states
for the even-even nuclei under consideration. 
}
\label{fig:wf}
\end{center}
\end{figure}

Figure~\ref{fig:energy} shows
the IBMCM energy spectra
and those in the standard IBM-2
without configuration mixing
\cite{nomura2025bb}.
The configuration mixing
lowers the $0^+_2$ energy-levels
in $^{76}$Se.
The mixing does not
change much the excitation
energies for $^{96}$Zr in the RHB-IBMCM.
This is because the local
prolate minimum appears
at an energy higher than
the global minimum by $\approx 2$ MeV,
and does not play a significant role.
The HFB-IBMCM reproduces
rather well the observed
$2^+_1$, $0^+_2$, and $2^+_2$
in $^{96}$Zr.
The quantitative differences between
the RHB- and HFB-IBMCM
results for $^{96}$Zr
are attributed to the fact that
the spherical global minimum
is obtained in the HFB PES,
while the pronounced deformation
is suggested in the RHB PES.
For $^{100}$Mo, with the configuration
mixing the $0^+_2$ energy-level
is considerably lowered to be
consistent with the experimental data
in the case of the RHB.
For $^{116}$Cd,
the $0^+_2$ energy-level is lowered
by the configuration mixing
with the RHB input.
The configuration mixing does not
affect the energy
spectra in $^{116}$Sn in either
case of the RHB and HFB,
since a prolate local minimum
found in this nucleus is
shown \cite{nomura2022bb,nomura2025bb}
to be at much higher energy than
the spherical equilibrium minimum
and is supposed to play
a negligible role.
Regarding $^{150}$Nd and $^{150}$Sm, 
the mapped IBMCM results show
much lower $0^+_2$ and $2^+_2$
levels than those obtained from
the IBM without
configuration mixing.

Figure~\ref{fig:wf} exhibits contributions
of the intruder
components to the IBMCM wave functions
of the $0^+_1$ and $0^+_2$ states.
The two configurations are
strongly mixed in the $0^+_1$ and $0^+_2$
wave functions for $^{76}$Se.
The intruder (oblate deformed)
configuration makes a minor contribution
to the $0^+_1$ ground state for $^{96}$Zr
in both the RHB and HFB,
but dominates the $0^+_2$ state
in the case of the HFB.
The $0^+_2$ wave functions for $^{100}$Mo 
and $^{116}$Cd are predominantly
composed of the intruder components
in the case of the RHB input,
and this structure accounts for,
to a large extent, the lowering 
of the $0^+_2$ levels in these nuclei.
Both the $0^+_1$ and
$0^+_2$ wave functions for $^{116}$Sn
are almost purely of normal (spherical)
configuration,
indicating that almost no mixing occurs.
For $^{150}$Nd and $^{150}$Sm,
certain amounts of the 
intruder configurations are present
in both the $0^+_1$ and $0^+_2$ states.

%
%
\begin{figure}[ht]
\begin{center}
\includegraphics[width=\linewidth]{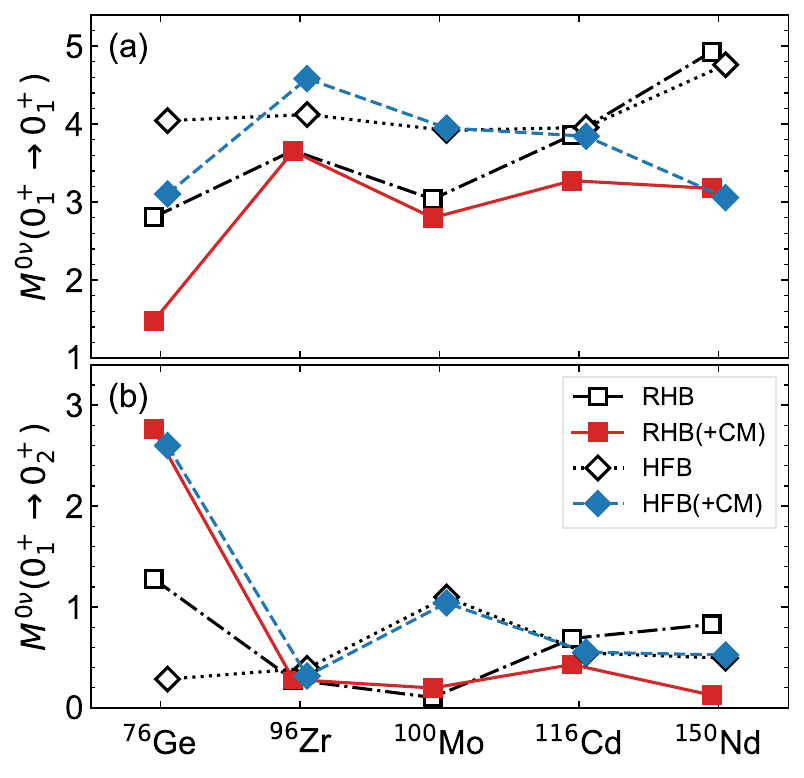}
\caption{
Predicted (a) $\mzn(0^+_1\,\to\,0^+_1)$
and (b) $\mzn(0^+_1\,\to\,0^+_2)$
for the studied even-even nuclei,
computed by the mapped IBM with and without
inclusion of the configuration mixing (CM).
}
\label{fig:nme}
\end{center}
\end{figure}

Figure~\ref{fig:nme} shows the predicted
NMEs $\mzn$ obtained from the
IBMCM for the
$0^+_{1}$ $\to$ $0^+_{1}$ and
$0^+_{1}$ $\to$ $0^+_{2}$
$\znbb$ decays of interest,
and those of the previous study
\cite{nomura2025bb}
using the IBM without
configuration mixing.
Tables~\ref{tab:nme1} and \ref{tab:nme2}
summarize the $\mgt$, $\mfe$, $\mte$,
and final $\mzn$ values obtained
from the mapped IBMCM,
and those $\mzn$ values
of \cite{nomura2025bb},
denoted $\mzn_{\rm w/o\,CM}$,
for comparisons.

The NMEs $\mzn(0^+_1\,\to\,0^+_1)$
for $^{76}$Ge are significantly reduced
by the spherical-oblate configuration
mixing in the daughter nucleus $^{76}$Se.
By the configuration mixing,
the $\mzn(0^+_1\,\to\,0^+_2)$
values for $^{76}$Ge are enhanced.
The present RHB-IBMCM result of
$\mzn(0^+_1\,\to\,0^+_1)$
for $^{76}$Ge is rather small compared
with majority of other theoretical values,
but is within the error bar of
the value predicted by the
In-Medium Similarity
Renormalization Group (IMSRG) method,
$2.60^{+128}_{-136}$ \cite{belley2024}, 
and is also close to those obtained
from the Effective Field Theory
\cite{brase2022}, which gave the
minimal
$0.63^{+86}_{-42}$ and
maximal
$0.95^{+146}_{-67}$ values.
The HFB-IBMCM NME for
the same decay is also close to the
IMSRG value of \cite{belley2024},
and to the recent NSM ones:
2.66 \cite{corragio2020},
2.72-4.38 \cite{jokiniemi2023}, and
$3.57\pm0.25$ \cite{castillo2025}.
It is also noted that the
present $\mzn(0^+_1\,\to\,0^+_2)$
are close to
the earlier IBM-2 value
of 2.02 \cite{barea2015}.

The configuration mixing
does not play a role
for the $^{96}$Zr $\znbb$ decay
with the RHB, as the local
prolate minimum in the PES is minor.
The HFB-IBMCM value of
$\mzn(0^+_1\,\to\,0^+_1)$ for $^{96}$Zr
is greater than that in
the previous IBM-2
mapping calculation \cite{nomura2025bb},
which took into account
the single spherical minimum.
In this particular case,
the single $d$-boson
energy for the unperturbed IBM-2
Hamiltonian for the spherical
normal configuration in the present
calculation for $^{96}$Zr
is larger than that in
\cite{nomura2025bb} by a
factor of 2.4.
Taking a large $d$-boson energy
generally enhances the
monopole contribution
to $\mzn$, hence a large NME.
The $\mzn(0^+_1\,\to\,0^+_1)$
values for the $^{96}$Zr decay
with both the RHB and HFB inputs
are more or less similar to
the predicted values from the QRPA
(within the range $2.4 \sim 5$)
\cite{simkovic2018,hyvarinen2015,jokiniemi2023}
and IBM-2 (3.92) \cite{deppisch2020}.

%
%
\begin{table}[ht]
\caption{\label{tab:nme1}
Predicted GT $\mgt$, Fermi $\mfe$,
tensor $\mte$, and final $\mzn$ matrix elements
for the $0^+_1$ $\to$ $0^+_1$ decays
of interest.
The $\mzn_{\rm w/o\,CM}$
values in the sixth column
denote those $\mzn$ values
of \cite{nomura2025bb},
without configuration mixing.
Entries in the upper and lower rows
shown for each decay process correspond
to the results from the
RHB- and HFB-IBMCM, respectively.
}
 \begin{center}
  \begin{tabular}{lccccc}
\hline
Decay & $\mgt$ & $\mfe$ & $\mte$ & $\mzn$ & $\mzn_{\rm w/o\,CM}$ \\
\hline
\multirow{2}{*}{$^{76}$Ge $\to$ $^{76}$Se} & $1.48$ & $-0.06$ & $-0.04$ & $1.48$ & $2.81$ \\ & $3.12$ & $-0.13$ & $-0.09$ & $3.10$ & $4.05$ \\
[1.0ex]
\multirow{2}{*}{$^{96}$Zr $\to$ $^{96}$Mo} & $3.25$ & $-0.42$ & $0.14$ & $3.65$ & $3.66$ \\ & $4.04$ & $-0.64$ & $0.15$ & $4.58$ & $4.12$ \\
[1.0ex]
\multirow{2}{*}{$^{100}$Mo $\to$ $^{100}$Ru} & $2.51$ & $-0.26$ & $0.12$ & $2.80$ & $3.04$ \\ & $3.52$ & $-0.45$ & $0.16$ & $3.95$ & $3.92$ \\
[1.0ex]
\multirow{2}{*}{$^{116}$Cd $\to$ $^{116}$Sn} & $2.87$ & $-0.48$ & $0.11$ & $3.28$ & $3.86$ \\ & $3.38$ & $-0.53$ & $0.13$ & $3.85$ & $3.96$ \\
[1.0ex]
\multirow{2}{*}{$^{150}$Nd $\to$ $^{150}$Sm} & $2.75$ & $-0.54$ & $0.09$ & $3.18$ & $4.93$ \\ & $2.62$ & $-0.60$ & $0.07$ & $3.06$ & $4.76$ \\
\hline
  \end{tabular}
 \end{center}
\end{table}

%
%
\begin{table}[ht]
\caption{\label{tab:nme2}
Same as the caption to Table~\ref{tab:nme1},
but for the $0^+_1$ $\to$ $0^+_2$ decays.
}
 \begin{center}
  \begin{tabular}{lccccc}
\hline
Decay & $\mgt$ & $\mfe$ & $\mte$ & $\mzn$ & $\mzn_{\rm w/o\,CM}$ \\
\hline
\multirow{2}{*}{$^{76}$Ge $\to$ $^{76}$Se} & $2.77$ & $-0.11$ & $-0.07$ & $2.76$ & $1.28$ \\ & $2.62$ & $-0.11$ & $-0.08$ & $2.60$ & $0.29$ \\
[1.0ex]
\multirow{2}{*}{$^{96}$Zr $\to$ $^{96}$Mo} & $0.25$ & $-0.04$ & $0.01$ & $0.28$ & $0.28$ \\ & $0.28$ & $-0.04$ & $0.01$ & $0.32$ & $0.39$ \\
[1.0ex]
\multirow{2}{*}{$^{100}$Mo $\to$ $^{100}$Ru} & $0.18$ & $-0.02$ & $0.01$ & $0.20$ & $0.10$ \\ & $0.93$ & $-0.11$ & $0.04$ & $1.04$ & $1.10$ \\
[1.0ex]
\multirow{2}{*}{$^{116}$Cd $\to$ $^{116}$Sn} & $0.38$ & $-0.07$ & $0.01$ & $0.43$ & $0.69$ \\ & $0.49$ & $-0.07$ & $0.02$ & $0.55$ & $0.55$ \\
[1.0ex]
\multirow{2}{*}{$^{150}$Nd $\to$ $^{150}$Sm} & $0.11$ & $-0.02$ & $0.01$ & $0.13$ & $0.83$ \\ & $0.46$ & $-0.07$ & $0.02$ & $0.53$ & $0.49$ \\
\hline
  \end{tabular}
 \end{center}
\end{table}

The $\mzn(0^+_1\,\to\,0^+_{1})$ value
for $^{100}$Mo in the case of
the RHB input
is also reduced as a result of
a strong mixing of the triaxial and large
prolate deformed configurations in
the $0^+$ wave functions
(cf. Fig.~\ref{fig:wf}).
The HFB-IBMCM suggests
for $^{100}$Mo that effects of the
configuration mixing are negligible
in the calculations of
$\mzn(0^+_1\,\to\,0^+_{1,2})$,
since essentially
no mixing occurs in the
$0^+_1$ and $0^+_2$ wave functions
(see Fig.~\ref{fig:wf}).
For $^{100}$Mo the HFB-IBMCM gives
the $\mzn(0^+_1\,\to\,0^+_1)$
value close to those
of the QRPA (3.90) \cite{hyvarinen2015},
and IBM-2 (3.84) \cite{barea2015}.
The RHB-IBMCM result for the
same decay is closer to the 
NSM value, 2.24 \cite{corragio2020}.
The present HFB result for
$\mzn(0^+_1\,\to\,0^+_2)$ for the
$^{100}$Mo decay is close to
the IBM-2 value of 1.12 \cite{barea2015}.

The $\mzn(0^+_1\,\to\,0^+_{1,2})$
NME values
for $^{116}$Cd are reduced by
the configuration mixing
by $\approx$ 15\% in the case of the RHB.
The configuration mixing is
included for the daughter
nucleus $^{116}$Sn in both the
RHB and HFB cases, but the mixing
effect in this nucleus
was shown to be quite minor
in Fig.~\ref{fig:wf}.
The reduction of the NMEs for
the $^{116}$Cd decay with the
RHB input is, therefore, mainly
due to the spherical-oblate
configuration mixing in the
parent nucleus $^{116}$Cd.
The NME $\mzn(0^+_1\,\to\,0^+_{1})$
of the RHB-IBMCM
for the $^{116}$Cd decay
is within the range of values
of the IBM-2
(2.83 \cite{barea2015}
and 2.98 \cite{deppisch2020}),
and of the QRPA
(3.74 \cite{simkovic2018},
4.26 \cite{hyvarinen2015},
and 2.93-5.70 \cite{jokiniemi2023}).

The $\mzn(0^+_1\,\to\,0^+_{1})$ values
for the $^{150}$Nd decay are
significantly reduced by
the inclusion of the configuration mixing
by $\approx$ 35\%.
With the RHB, the
$\mzn(0^+_1\,\to\,0^+_{2})$
value for $^{150}$Nd is also reduced
by the shape mixing.
The mapped IBM calculation
without the configuration mixing
\cite{nomura2025bb}
provided the
$\mzn(0^+_1\,\to\,0^+_{1})$ values
for the $^{150}$Nd decay that are
substantially large.
The present IBMCM, however,
gives the NMEs that are rather close
to those from the QRPA, e.g.,
2.71 \cite{mustonen2013}, 3.01 \cite{fang2018},
and 3.85 \cite{terasaki2020},
and those from the IBM-2,
2.47 \cite{barea2015} and
3.57 \cite{deppisch2020}.

%
%
\begin{figure}[ht]
\begin{center}
\includegraphics[width=\linewidth]{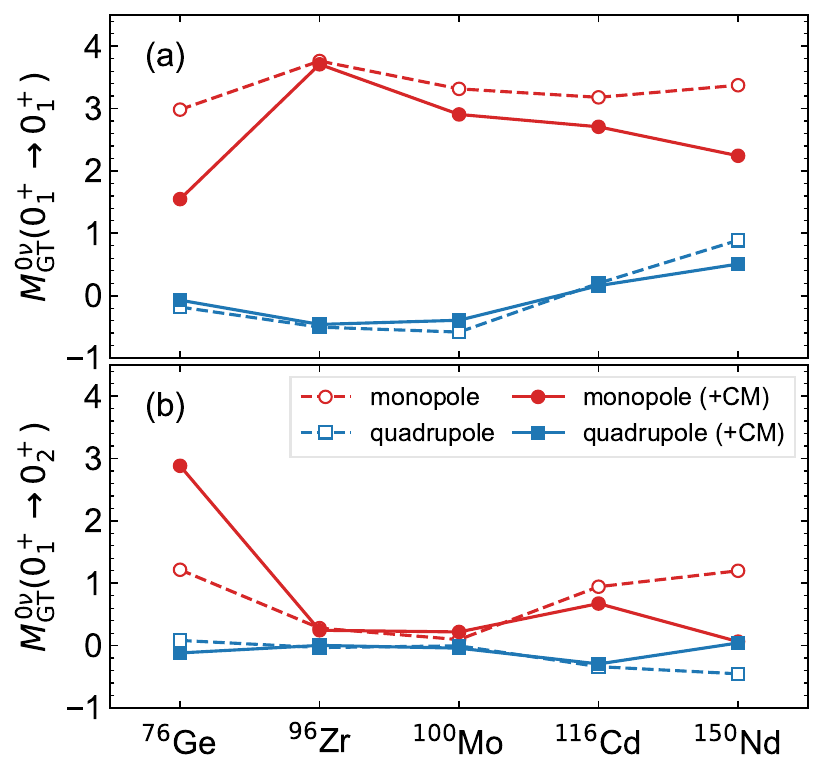}
\caption{Decomposition of 
(a) $\mgt(0^+_1\,\to\,0^{+}_1)$
and (b) $\mgt(0^+_1\,\to\,0^{+}_2)$
into the monopole and 
quadrupole components calculated
with and without the configuration mixing.
Note that in (b)
the sign of the monopole
and quadrupole parts
of $\mgt(0^+_1\,\to\,0^{+}_2)$
is set so that the
former is positive.
The RHB-SCMF calculation is
performed to provide the
microscopic input.}
\label{fig:ssdd}
\end{center}
\end{figure}

%
%
\begin{figure}[ht]
\begin{center}
\includegraphics[width=\linewidth]{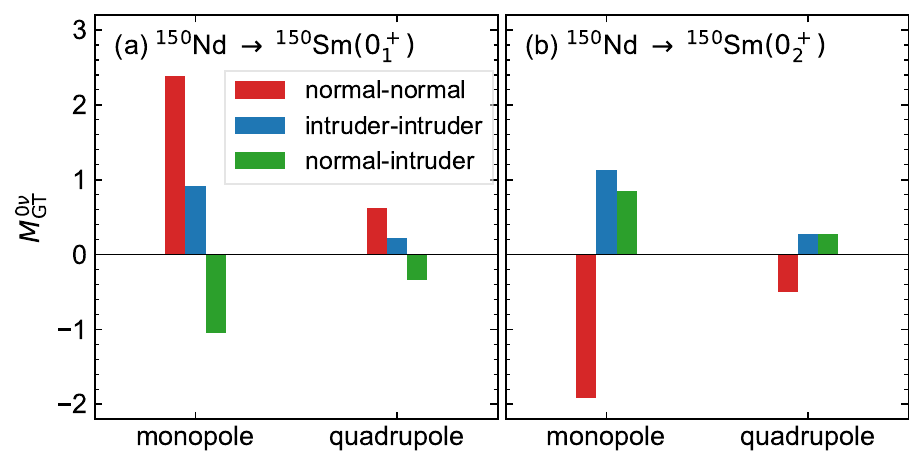}
\caption{Decomposition
of monopole and quadrupole parts
of $\mgt$ for the $\znbb$ decays
(a) $^{150}$Nd$(0^+_1)\,\to\,^{150}$Sm$(0^+_1)$
and
(b) $^{150}$Nd$(0^+_1)\,\to\,^{150}$Sm$(0^+_2)$
into the normal-to-normal
$[N_{\nu,{\rm I}}\otimes N_{\pi,{\rm I}}]\,\to\,
[N_{\nu,{\rm F}}\otimes N_{\pi,{\rm F}}]$,
intruder-to-intruder
$[N_{\nu,{\rm I}}\otimes N_{\pi,{\rm I}}']\,\to\,
[N_{\nu,{\rm F}}\otimes N_{\pi,{\rm F}}']$,
and
normal-to-intruder
$[N_{\nu,{\rm I}}\otimes N_{\pi,{\rm I}}]\,\to\,
[N_{\nu,{\rm F}}\otimes N_{\pi,{\rm F}}']$
components of the one-boson transfer operator.
The RHB-SCMF calculation is
performed to provide the
microscopic input.
}
\label{fig:decom-pc}
\end{center}
\end{figure}

Figure~\ref{fig:ssdd} gives
contributions of the
monopole and quadrupole components
in the calculated $\mgt$
with the RHB input.
Note a similar argument
holds for the HFB input.
The configuration mixing generally reduces
the monopole parts of $\mgt(0^+_1\,\to\,0^+_{1})$,
and also reduces the
quadrupole parts slightly.
The suppression of the monopole
contribution reflects that the
$s$ boson content in
the ground states is reduced due to
the inclusion of deformed
intruder configurations.
It is, however, rather difficult
to draw a general conclusion
concerning the
configuration mixing effect
on the $0^+_1\,\to\,0^+_{2}$
decay.
For instance,
the monopole component of
$\mgt(0^+_1\,\to\,0^+_{2})$ for
the $^{76}$Ge decay is
increased by the configuration
mixing.
It appears that the
$\mzn(0^+_1\,\to\,0^+_{2})$ values
are rather sensitive to the structure
of the $0^+_2$ wave functions,
which may also depend on
some missing correlations.

For the $^{116}$Cd and $^{150}$Nd
decays in particular,
the one-boson
transfers between the intruder and intruder
and between the normal and intruder
configurations contribute to the NMEs,
since the configuration mixing
is performed for both
the parent and daughter nuclei
for these decay processes.
Figure~\ref{fig:decom-pc} displays
decomposition of $\mgt$
for the $^{150}$Nd decay
into the
$[N_{\nu,{\rm I}}\otimes N_{\pi,{\rm I}}]\,\to\,
[N_{\nu,{\rm F}}\otimes N_{\pi,{\rm F}}]$, 
$[N_{\nu,{\rm I}}\otimes N_{\pi,{\rm I}}']\,\to\,
[N_{\nu,{\rm F}}\otimes N_{\pi,{\rm F}}']$,
and
$[N_{\nu,{\rm I}}\otimes N_{\pi,{\rm I}}]\,\to\,
[N_{\nu,{\rm F}}\otimes N_{\pi,{\rm F}}']$
components in
the monopole and quadrupole parts.
For the $0^+_1\,\to\,0^+_1$ decay, the
$[N_{\nu,{\rm I}}\otimes N_{\pi,{\rm I}}']\,\to\,
[N_{\nu,{\rm F}}\otimes N_{\pi,{\rm F}}']$,
and
$[N_{\nu,{\rm I}}\otimes N_{\pi,{\rm I}}]\,\to\,
[N_{\nu,{\rm F}}\otimes N_{\pi,{\rm F}}']$
components make sizable contributions
to the monopole and quadruple parts,
but have opposite sign
to approximately cancel.
For the $0^+_1\,\to\,0^+_2$ decay,
these two components contribute
coherently to $\mgt$
and cancel the contribution of
the $[N_{\nu,{\rm I}}\otimes N_{\pi,{\rm I}}]\,\to\,
[N_{\nu,{\rm F}}\otimes N_{\pi,{\rm F}}]$ component.
This explains the very small
$\mzn(0^+_1\,\to\,0^+_2)$ values
for the $^{150}$Nd decay,
particularly in the RHB-IBMCM.
For the $^{116}$Cd,
since the intruder configurations
do not play a significant role
in $^{116}$Sn (cf. Fig.~\ref{fig:wf}),
contributions of the
intruder-to-intruder and
normal-to-intruder transfers
to the NME are negligible.

To summarize,
the mixing of multiple shape
configurations has been
incorporated in the
predictions of $\znbb$-decay NMEs
using the IBM with configuration mixing,
which is formulated
by the SCMF methods.
Several even-even nuclei that are
$\znbb$ decay emitters or/and final-state
nuclei are characterized by the presence
of multiple minima in the energy surfaces.
The $\znbb$ decay operators
have been extended to
allow transitions between
the intruder and intruder
and between the normal
and intruder configurations.
It has been shown that
the shape mixing
occurs in the ground
and excited $0^+$ states
in the studied even-even nuclei.
The inclusion of the configuration mixing
improves an overall description of
the low-energy spectra,
in particular,
that of the $0^+_2$ state.
The configuration mixing
substantially changes the monopole
part of the NME, and generally
reduces the values of the
$0^+_1$ $\to$ $0^+_1$ $\znbb$ decay NMEs.
The result of this work shows
relevance of the shape coexistence and
mixing in the $\znbb$-decay NMEs predictions.

\section*{Acknowledgements}

This work has been supported
by JSPS KAKENHI Grant No. JP25K07293.


\bibliographystyle{elsarticle-num} 
\bibliography{refs}

\end{document}